\documentclass[amsmath,amssymb,nofootinbib,notitlepage,superscriptaddress,twocolumn]{revtex4-2}

\usepackage{comment}
\usepackage{amsmath,amsfonts,amsthm,bm} 
\usepackage{graphicx}
\usepackage{dcolumn}
\usepackage{bm}
\usepackage{color}
\definecolor{RedWine}{rgb}{0.743,0,0}
\definecolor{RoyalBlue}{rgb}{0.25,.41,.88}
\definecolor{ForestGreen}{rgb}{.13,.54,.13}
\usepackage[backref,breaklinks,colorlinks,citecolor=ForestGreen]{hyperref}

\begin{document}

\preprint{APS/123-QED}

\title{Dynamics of ultrarelativistic charged particles with strong radiation reaction. \\  II. Entry into Aristotelian equilibrium}

\author{Yangyang Cai}
\author{Samuel E. Gralla}
\affiliation{Department of Physics, University of Arizona, 1118 E 4th Street, U.S.A}
\author{Vasileios Paschalidis}
\affiliation{Department of Physics, University of Arizona, 1118 E 4th Street, U.S.A}
 \affiliation{Department of Astronomy, University of Arizona, 933 N Cherry Ave, U.S.A}

\begin{abstract}
As first proposed by Gruzinov, a charged particle moving in strong electromagnetic fields can enter an equilibrium state where the power input from the electric field is balanced by radiative losses.  When this occurs, the particle moves at nearly light speed along special directions called the principal null directions (PNDs) of the electromagnetic field.  This equilibrium is  ``Aristotelian'' in that the particle velocity, rather than acceleration, is determined by the local electromagnetic field.  In paper I of this series, we analytically derived the complete formula for the particle velocity at leading order in its deviation from the PND, starting from the fundamental Landau-Lifshitz (LL) equation governing charged particle motion, and demonstrated agreement with numerical solutions of the LL equation.  We also identified five necessary conditions on the field configuration for the equilibrium to occur.  In this paper we study the entry into equilibrium using a similar combination of analytical and numerical techniques.  We simplify the necessary conditions and provide strong numerical evidence that they are also sufficient for equilibrium to occur.  Based on exact and approximate solutions to the LL equation, we identify key timescales and properties of entry into equilibrium and show quantitative agreement with numerical simulations.  Part of this analysis shows analytically that the equilibrium is linearly stable and identifies the presence of oscillations during entry, which may have distinctive radiative signatures.  Our results provide a solid foundation for using the Aristotelian approximation when modeling relativistic plasmas with strong electromagnetic fields.
\end{abstract}

\maketitle

\section{Introduction}

Every electromagnetic field defines, algebraically at each point, a pair of (possibly identical) light-speed velocities, its principal null directions (PNDs) \cite{synge1956relativity}.  Recently, it has become clear that this mathematical notion has an elegant physical manifestation: \textit{ultrarelativistic charged particles follow the PNDs}.  The phenomenon appears to be quite universal in that it emerges in different regimes and with different mechanisms regulating the ultimate particle speed, such as classical radiation reaction in magnetically dominated fields relevant to astrophysics \cite{mestel1985,BFink1989,gruzinov2013b,Jacobson:2015cia,Petri:2019tix,Cao:2019uhv} or quantum radiation reaction in nearly null fields relevant to laser-plasma physics \cite{Gonoskov:2017lyz,Samsonov:2018skj,Ekman:2021vwg,Gonoskov:2021hwf,Samsonov2022}.  This regime is \textit{Aristotelian} in that the particle velocity, rather than acceleration, is determined by the local electromagnetic field.

In paper I of this series \cite{Cai:2022mkw} we initiated a detailed study of Aristotelian motion for classical charged particles in strong external fields.  We considered the fundamental Landau-Lifshitz (LL) equation \cite{LL}, which includes both Lorentz force and self-force.  We adopted the approximation that a particle is nearly, but not exactly, moving on a PND, and derived equations for its velocity at leading order in the deviation from the PND.  We identified precise conditions on the field configuration that are necessary for the equilibrium to occur.  Finally, we demonstrated numerical agreement of this approximation with full solutions of the LL equation in the appropriate regime, using a new numerical code.

In this paper we will use similar analytical and numerical techniques to study the entry  into Aristotelian equilibrium.  Our main results are: (1) the necessary conditions identified in paper I are in fact sufficient for equilibrium to occur; (2) the equilibrium is linearly stable; (3) in some parameter ranges there are oscillations during the approach to equilibrium, whose properties we study analytically.  Together with the findings of paper I, these results provide a definite prescription for using the Aristotelian approximation in practice.

This paper is organized as follows.  In Sec.~\ref{sec:equilibrium} we review the Aristotelian equilibrium and relate the assumptions and results of paper I \cite{Cai:2022mkw} to the simple versions given originally by Gruzinov \cite{Gruzinov:2013pva,gruzinov2013b}.  In Sec.~\ref{sec:numerical} we review the LL equation and introduce notation.  In Sec.~\ref{sec:torsion} we show an example of Aristotelian equilibrium in a helical field configuration.  In Sec.~\ref{sec:approach} we analytically study the approach to equilibrium and demonstrate agreement with numerical simulations.  In Sec.~\ref{sec:C4} we perform a large numerical parameter survey that validates our conditions for entry into equilibrium.  Finally, in Sec.~\ref{sec:summary} we summarize our results, focusing on a simple prescription for using the Aristotelian approximation in astronomical modeling.  Appendix~\ref{sec:numerical-details} provides details of our numerical scheme.  We use Gaussian units with the speed of light set equal to one.

\section{Aristotelian Equilibrium}\label{sec:equilibrium}

In this section we review the properties of the PNDs  and the Aristotelian equilibrium.  The PNDs \cite{synge1956relativity} $\ell_+^\mu$ and $\ell_-^\mu$ of an electromagnetic field $F_{\mu \nu}$ are the solutions to the pointwise eigenvalue equation 
\begin{align}
    F^{\mu}{}_{\nu}\ell_{\pm}^\nu = \pm E_0 \ell_{\pm}^\nu.
\end{align}
The explicit solution in terms of electric and magnetic fields is
\begin{align}
\ell^\mu_{\pm} & = (1,\Vec{v}_\pm), \label{v} \\
\Vec{v}_{\pm} & =\frac{\Vec{E}\times\Vec{B}\pm(B_0\Vec{B}+E_0\Vec{E})}{B^2+E_0^2}, \label{vpm}
\end{align} 
where $E_0$ and $B_0$ are given in terms of the invariants $P=\Vec{B}^2-\Vec{E}^2$ and $Q=\Vec{E}\cdot\Vec{B}$ as 
\begin{align}
    E_0 & = \sqrt{\sqrt{(P/2)^2+Q^2}-P/2} \label{E0} \\
    B_0 & = \textrm{sign}(Q) \sqrt{\sqrt{(P/2)^2+Q^2}+P/2}. \label{B0}
\end{align}
When the PNDs are degenerate ($\ell^\mu_+=\ell^\mu_-$), the eigenvalue vanishes and the field is null ($E_0=B_0=0$), in which case the electric and magnetic fields are orthogonal and equal in magnitude in any Lorentz frame.  When the PNDs are distinct, the eigenvalues are $\pm E_0$, and $E_0>0$ is the magnitude of the electric field in any frame where the electric and magnetic fields are parallel.  Similarly, $|B_0|$ is the magnitude of the magnetic field in such a frame, with $B_0$ positive/negative when the fields are aligned/antialigned.

The PNDs define integral curves $\vec{x}_\pm(t)$ by the equation  $d\vec{x}_\pm/dt=v_\pm$.  The parameter $t$ is the arc length of the space curve, since $v_\pm^2=1$.  On each curve we may erect a Frenet Serret frame $\{\vec{\ell},\vec{n},\vec{k}\}$ with $\vec{\ell}=v_\pm$ \cite{Cai:2022mkw}.  Since these curves fill space, for each choice of $\pm$ we have a full orthonormal basis for vector fields.  In particular, we may decompose the velocity vector of a charged particle as
\begin{align}\label{vdecomp}
\Vec{v}=v_\ell\Vec{\ell}+v_n\Vec{n}+v_k\Vec{k},
\end{align}
where we choose $\vec{\ell}=\vec{v}_+$ for positively charged particles and $\vec{\ell}=\vec{v}_-$ for negatively charged particles.  As this is an orthonormal frame, the Lorentz factor is reconstructed by
\begin{align}\label{gammafun}
    \gamma = \frac{1}{\sqrt{1-v_\ell^2-v_n^2-v_k^2}}.
\end{align}

The Frenet-Serret vectors $\{ \vec{\ell}, \vec{n}, \vec{k} \}$ and their associated curvature $\kappa$ and torsion $\iota$ are local functions of the electric and magnetic fields for each choice of $\pm$.  This may be seen from the defining equations,
\begin{align}
    \vec{\ell} & = \vec{v}_{\pm}, \label{fs1} \\
    \kappa \vec{n} & =(\vec{\ell} \cdot \vec{\nabla}) \vec{\ell}, \qquad (\vec{n} \cdot \vec{n} = 1) \label{fs2} \\
    \vec{k} & = \vec{\ell} \times \vec{n}, \label{fs3} \\
    \iota \vec{n} & = - (\vec{\ell} \cdot \vec{\nabla}) \vec{k}. \label{fs4}
\end{align}
The first vector $\vec{\ell}=v_{\pm}$ is determined by the values of $\vec{E}$ and $\vec{B}$ via Eqs.~\eqref{fs1} and \eqref{vpm}.  The second vector $\hat{n}$ and the curvature $\kappa$ involve first derivatives as well  [Eq.~\eqref{fs2}].  The third vector $\vec{k} = \vec{\ell} \times \vec{n}$ also depends on first derivatives.  Finally, the torsion $\iota$ depends on first and second derivatives [Eq.~\eqref{fs3}].  We will also define the radius of curvature $R$,
\begin{align}
    R = \frac{1}{\kappa}.
\end{align}

Note that while $E_0$, $B_0$ and $\ell^\mu$ are invariant notions, the projection of the null vector $\ell^\mu$ to the spatial velocity $v_{\pm}=\vec{\ell}$ depends on the choice of Lorentz frame.  The corresponding Frenet-Serret basis, together with its associated curvature and torsion, are similarly non-invariant.  We will be formulating assumptions and deriving results in terms of these non-invariant quantities; the interpretation is that our results hold in frames satisfying our assumptions.  Note that we will use the phrase ``PND'' for both the invariant null direction in a spacetime sense and the non-invariant spatial direction $\vec{v}_\pm$ in a given frame.  Context will make clear which notion is meant.

A particle of charge $q$ and mass $m$ defines a length scale $\mathcal{R}$ and a field scale $\mathcal{E}$ by
\begin{align}\label{scales}
    \mathcal{R} \equiv \frac{q^2}{m}, \qquad \mathcal{E} \equiv \frac{3}{2} \frac{m^2}{|q|^3},
\end{align}
with a conventional factor of $3/2$. The ``classical electron radius'' $\mathcal{R}$ is the distance where the electrostatic self-energy of a point charge equals its rest mass, and the ``classical critical field'' $\mathcal{E}$ is (three-halves times) the strength of the electric field at this location.  These represent typical scales at which the classical description of the particle will break down.  These quantities also define time and magnetic field scales after multiplying by suitable factors of the speed of light (set here to unity).  We therefore assume
\begin{align}\label{classicalok}
    E_0,B_0 \ll \mathcal{E} \qquad L,T \gg \mathcal{R},
\end{align}
where $L$ and $T$ are typical length and time scales for the field configuration to change significantly.

\subsection{Original Derivation}\label{sec:original}

We now summarize Gruzinov's original arguments for Aristotelian motion \cite{gruzinov2013b,Gruzinov:2013pva} using the notation of our paper.  Gruzinov considered the case where a particle moves primarily along a PND, i.e.,
\begin{align}
    \gamma & \gg 1, \label{c1} \\
    \sqrt{v_n^2+v_k^2} & \ll 1. \label{c2}
\end{align}
If we approximate the particle motion as a circle of radius equal to the radius of curvature $R=1/\kappa$ of the PND, the power radiated is $(2/3)q^2 R^2 \gamma^4$.  Gruzinov assumed that this ``curvature radiation'' power is balanced by the Lorentz force power $|q| E_0$,
\begin{align}\label{power-balance}
    \frac{2}{3} q^2 R^2 \gamma^4 = |q| E_0.
\end{align}
Solving for $\gamma$ gives the Gruzinov formula for the Lorentz factor,
\begin{align}\label{gruzigamma}
    \gamma_g & = \left( \frac{3E_0 R^2}{2|q|} \right)^{1/4}.
\end{align}

Gruzinov obtained an additional condition for this equilibrium based on the idea that it can only occur when the particle has enough space to be accelerated to this terminal Lorentz factor before the PND curves significantly.  The curvature radius $R$ sets the scale over which the uniform field approximation breaks down, so the particle must be able to gain $m \gamma_g$ energy in a region much smaller than $R$.  The typical energy gain over a region of size $D$ is $q E_0 D$, so we obtain the condition
\begin{align}
    m \gamma_g \ll q E_0 R,
\end{align}
which may equivalently be written
\begin{align}
\gamma_g \ll \frac{E_0 R}{\mathcal{E} \mathcal{R}}
\end{align}
or
\begin{align}\label{gruzicond}
\mathcal{E}^{3/2} \mathcal{R} \ll E_0^{3/2} R.
\end{align}

\subsection{LL Derivation}

In paper I \cite{Cai:2022mkw} we sought to better understand the Aristotelian equilibrium by studying the fundamental LL equation of charged particle dynamics.  We again assumed motion along a PND [Eqs.~\eqref{c1} and \eqref{c2}].  However, the energy balance condition \eqref{power-balance} is inappropriate in this context since (1) it requires the particle motion to be treated as circular, an uncontrolled approximation whose compatibility with the LL equation is not obvious; and (2) the local LL dynamics contains more information than just energy conservation.  Instead, to express the idea of energy balance, we assumed that the local change in energy is small compared to the typical value set by the Lorentz force,
\begin{align}
    m \left|\frac{d\gamma}{dt}\right| & \ll |q| E_0. \label{c4} 
\end{align}
We also assumed that the timescale $T$ for changes in the field is long compared to the lengthscale $L$ for spatial changes in the field,
\begin{align}\label{c5}
    T \gg L.
\end{align}
We found that Gruzinov's equilibrium emerged only after a final additional assumption,
\begin{align}
    |\iota| & \ll \frac{|q|}{m\gamma}\textrm{Max}\{E_0,|B_0|\} ,\label{c3}
\end{align}
where $\iota$ is the torsion of the PND.  These three conditions \eqref{c4}, \eqref{c5} and \eqref{c3} replace the assumption \eqref{power-balance} of global power balance in approximate circular motion.  Eq.~\eqref{c4} guarantees approximate power balance locally at the level of the LL equation (as opposed to globally, at the level of total radiated energy), while Eqs.~\eqref{c5} and \eqref{c4} reflect the approximate circular motion with radius $R$.

Under these five assumptions \eqref{c1}, \eqref{c2}, \eqref{c4}, \eqref{c5}, \eqref{c3}, we found that the field determines the velocities pointwise as
\begin{align}
    \gamma&=\left( \frac{9}{4}\frac{R^2}{\mathcal{R}^2}\frac{E_0}{\mathcal{E}} \right)^{1/4} = \gamma_g \label{gammafinal} \\
    v_n&=-\frac{1+\delta}{\gamma}\sqrt{\frac{E_0}{\mathcal{E}}} \label{vnfinal}\\
    v_k&=\frac{\delta}{\gamma}\frac{B_0}{E_0}\sqrt{\frac{E_0}{\mathcal{E}}},\label{vkfinal}
\end{align}
where we used the definitions \eqref{scales} and also introduced
\begin{align}\label{delta}
    \delta = \frac{E_0\mathcal{E}}{E_0^2+B_0^2}.
\end{align}
We thus reproduced Gruzinov's Lorentz factor \eqref{gruzigamma} and derived new expressions \eqref{vnfinal} and \eqref{vkfinal} for the drift velocities.

\subsection{Conditions on the field configuration}\label{sec:conditions}

The five assumptions \eqref{c1}, \eqref{c2}, and \eqref{c4}--\eqref{c3} can be recast as conditions purely on the field configuration by using the results \eqref{gammafinal}--\eqref{vkfinal}.  Presenting the results as five conditions $C_i\ll 1$, the conditions are
\begin{align}
    C_1 & = \frac{\mathcal{R}}{R} \sqrt{\frac{\mathcal{E}}{E_0}}  \ll 1 \label{C1} \\
    C_2 & = \delta \frac{\mathcal{R}}{R} \sqrt{\frac{\mathcal{E}}{E_0}} \ll 1 \label{C2} \\
    C_3 & = |\iota| \sqrt{R \mathcal{R}} \left(\frac{E_0}{\mathcal{E}}\right)^{1/4} \frac{\mathcal{E}}{\textrm{Max}\{E_0,|B_0|\}} \ll 1 \label{C3} \\
    C_4 & = \eta \sqrt{\frac{\mathcal{R}}{R}}\left(\frac{\mathcal{E}}{E_0}\right)^{3/4} \ll 1, \label{C4}\\
    C_5 &= \frac{L}{T} \ll 1 \label{C5}
\end{align}
with 
\begin{align}\label{eta}
   \eta = \left|\vec{v}\cdot \vec{\nabla} R +\frac{R}{2E_0}\vec{v}\cdot \vec{\nabla} E_0 \right|.
\end{align}
In this last expression, it is understood that Eqs.~\eqref{gammafinal}--\eqref{vkfinal} [as well as \eqref{gammafun}] are to be used to express $\vec{v}$ in terms of the local field configuration.  In obtaining \eqref{C2} we have used the fact that $\sqrt{v_n^2+v_k^2} \approx \sqrt{\delta}/\gamma$ under the assumptions \eqref{classicalok}.

The names $C_i$ are derived from paper I.  In this paper, $C_1$--$C_5$ correspond (respectively) to Eqs.~\eqref{c1}, \eqref{c2}, \eqref{c3}, \eqref{c4}, and \eqref{c5}.   In particular, $C_1 \ll 1$ and $C_2 \ll 1$ express the approximate PND motion \eqref{c1} and \eqref{c2}, while $C_4\ll1$ expresses the approximate local equilibrium \eqref{c4}.

Our expression for $C_4$ differs from that presented in paper I, where we assumed that $\eta \sim R/L$.  Generically we will have $\eta \sim 1$.  In this paper we also include cases where $\eta$ is very small, since they arise quite naturally in our numerical studies, and help to illustrate the differences between our approach and that originally given by Gruzinov (Sec.~\ref{sec:original}). 

Notice that $C_1$ and $C_2$ are related to $C_4$ by the equations
\begin{align}
\eta C_1^{3/2} & = \frac{\mathcal{R}}{R} C_4 \ll C_4 \label{ic41} \\
\eta C_2 & \leq \sqrt{\frac{\mathcal{R}}{R}} C_4 \ll C_4. \label{ic42}
\end{align}
In the second line we used the bound $\delta \leq \mathcal{E}/E_0$, which follows from the definition \eqref{delta} of $\delta$.  In the generic case that $\eta\sim 1$, we see that both $C_1$ and $C_2$ are small compared with $C_4$.  This means that $C_4 \ll 1$ in fact implies that $C_1\ll 1$ and $C_2 \ll 1$, and generically one can ignore the $C_1$ and $C_2$ conditions.  While $\eta$ can be a small number in special cases (such as the circular fields studied in Sec.~\ref{sec:circular} below), it would have to be less than $\sim\sqrt{\mathcal{R}/R}$ for the $C_1$ and $C_2$ conditions to become relevant.  For macroscopic fields this factor is very small; for example, $\sqrt{\mathcal{R}/R}\approx 5\times10^{-8}$ if $R$ is one meter.  A field configuration with $\eta$ this small would be extraordinarily fine-tuned.  We therefore conclude that for the macroscopic fields of relevance in astrophysics, we may always ignore the $C_1$ and $C_2$ conditions, since they are implied by the $C_4$ condition.

Let us therefore focus on the $C_4$ condition \eqref{C4}.  This condition can be rewritten as
\begin{align}\label{C4b}
\eta^2 \mathcal{R} \mathcal{E}^{3/2} \ll R E_0^{3/2},
\end{align}
showing that it is in fact equivalent to Gruzinov's condition \eqref{gruzicond} in the generic case $\eta\sim 1$.  This agreement is interesting since Gruzinov's condition \eqref{gruzicond} arose from reasoning about when a particle \textit{can} enter equilibrium, whereas our condition \eqref{C4b} arose from the assumption \eqref{c4} that equilibrium had been \textit{achieved}.  We will see numerically in Sec.~\ref{sec:C4} that Eq.~\eqref{C4b} is necessary and sufficient for equilibrium, provided the other conditions are satisfied.

Finally we discuss $C_3$, which satisfies
\begin{align}\label{C3b}
    C_3 & = \frac{|\iota|}{\kappa} \eta^{-1} C_4 \frac{E_0}{\textrm{Max}\{E_0,|B_0|\}}.
\end{align}
The last factor $E_0/\textrm{Max}\{E_0,|B_0|\}$ is always $\lesssim 1$ and will be small for pulsars, where $E_0 \ll |B_0|$.  Again assuming the generic case $\eta \sim 1$, Eq.~\eqref{C3b} shows that large values of the torsion-to-curvature ratio $|\iota|/\kappa$ are required for the condition $C_3 \ll 1$ to be violated in a regime where the condition $C_4 \ll 1$ is satisfied.  For fields with $|\iota|/\kappa \lesssim 1$, the $C_3$ condition can be ignored, since it is implied by the $C_4$ condition.  We discuss cases with large $|\iota|/\kappa$ in Sec.~\ref{sec:torsion} below.

The conditions $C_i \ll 1$ arose in paper I as a minimal set of assumptions under which the LL equation reduced to the simple results \eqref{gammafinal}--\eqref{vkfinal}.  Although we do not attempt any rigorous mathematical proof, it seems clear from the steps of the derivation there that these assumptions were all necessary for the simple results to emerge.  We therefore regard the conditions \eqref{C1}--\eqref{C5} as necessary conditions for the equilibrium described by \eqref{gammafinal}--\eqref{vkfinal}.  One of the main goals of this paper is to argue that they are also \textit{sufficient} for equilibrium to occur. 

\section{Landau-Lifshitz Equation}\label{sec:numerical}
The LL equation is reviewed in paper I \cite{Cai:2022mkw}; here we briefly recap the physics and introduce our notation for numerical simulations.  
In terms of $\vec{E}$ and $\vec{B}$ fields, the LL equation is\footnote{We have dropped terms involving the derivative of the field strength, which are always negligible in the regime of validity discussed below \eqref{fi}.}

\begin{align}
    \frac{d(\gamma\Vec{v})}{d\tau} = \frac{\gamma q}{m}\Bigg\{& \Vec{E}+\Vec{v}\times\Vec{B} \nonumber \\ 
    & \pm\frac{1}{\mathcal{E}}\left[(\Vec{E}\cdot\Vec{v})\Vec{E}+(\Vec{E}+\Vec{v}\times\Vec{B})\times\Vec{B}\right] \nonumber \\
    & \mp\frac{\gamma^2}{\mathcal{E}}\left[(\Vec{E}+\Vec{v}\times\Vec{B})^2-(\Vec{E}\cdot\Vec{v})^2\right]\Bigg\}, \label{fi}
\end{align}
where $\pm$ is the sign of $q$, $\vec{v}$ is the velocity, $\gamma=1/\sqrt{1-v^2}$ is the Lorentz factor, and $\tau$ is the proper time.  As discussed originally by LL \cite{LL}, this equation is valid provided that the corrections to Lorentz force motion (the final two terms) are small compared to the Lorentz force \textit{in the rest frame of the particle}.  However, in the lab frame they may be of comparable or greater magnitude, and this is the regime of interest to us.

For numerical purposes, we normalize the equation using the natural scales of the equation, together with a dimensionless number $\chi$ that can be chosen for convenience.  Defining
\begin{align}
    \Tilde{\tau}&=\frac{3}{2}\chi^2\frac{\tau}{\mathcal{R}}\\
    \Tilde{\Vec{E}}&=\frac{\Vec{E}}{\chi\mathcal{E}}\\
    \Tilde{\Vec{B}}&=\frac{\Vec{B}}{\chi\mathcal{E}}\\
    \Tilde{\Vec{p}}&=\frac{\Vec{p}}{m}=\gamma\Vec{v},
\end{align}
Eq.~\eqref{fi} becomes
\begin{align}\label{normeq}
    \frac{d\Tilde{\Vec{p}}}{d\Tilde{\tau}}&=\frac{1}{\chi}\Vec{f_L}\pm\left[(\Tilde{\Vec{E}}\cdot \Tilde{\Vec{p}})\Tilde{\Vec{E}}+\Vec{f_L}\times\Tilde{\Vec{B}}\right]\mp\left[f_L^2-(\Tilde{\Vec{E}}\cdot\Tilde{\Vec{p}})^2\right]\Tilde{\Vec{p}},
\end{align}
where
$\Vec{f_L}=\gamma\Tilde{\Vec{E}}+\Tilde{\Vec{p}}\times\Tilde{\Vec{B}}$ is the Lorentz force term.  The Lorentz factor is related to the rescaled momentum by
\begin{align}
    \gamma&=\sqrt{1+\Tilde{p}^2}.
\end{align}
Once the equation is solved for $\tilde{\vec{p}}(\tilde{\tau})$, the position can be recovered by a subsequent integration.  For convenience we define a normalized position vector $\Tilde{\Vec{x}}$,
\begin{align}
    \Tilde{\Vec{x}}&=\frac{3}{2}\chi^2\frac{\Vec{x}}{\mathcal{R}}=\frac{3}{2}\chi^2\frac{1}{\mathcal{R}}\int \Vec{v} dt=\int \Tilde{\Vec{p}}\ d\Tilde{\tau}.
\end{align}
If necessary, one can also determine the lab-frame time $t$ from $dt = \gamma d\tau$.

We will choose the value of $\chi$ so that lengths are measured in meters when the electron is considered,
\begin{align}
    \chi=\sqrt{\frac{2}{3}\frac{\mathcal{R}}{1 {\rm m}}}\approx 4.33\times 10^{-8}.
\end{align}
In particular for electrons we have
\begin{align}
    \tilde{\vec{x}} & = \textrm{(position in meters)} \label{xphys}\\
    \tilde{\tau} & = \textrm{(proper time in meters)} \label{tauphys}\\
    \tilde{\vec{E}} & = \textrm{(electric field in units of $10^{13}$ V/m)} \label{Ephys} \\
    \tilde{\vec{B}} & = \textrm{(magnetic field in units of $10^8$ G)}, \label{Bphys}
\end{align}
These are somewhat convenient for pulsars, which have radii of $\sim10$km and magnetic fields of $10^8$--$10^{15}$G.

Our numerical method is described in Appendix.~\ref{sec:numerical-details}.

\section{An example with torsion}\label{sec:torsion}

\begin{figure}
    \centering
    \includegraphics[width=.45\textwidth]{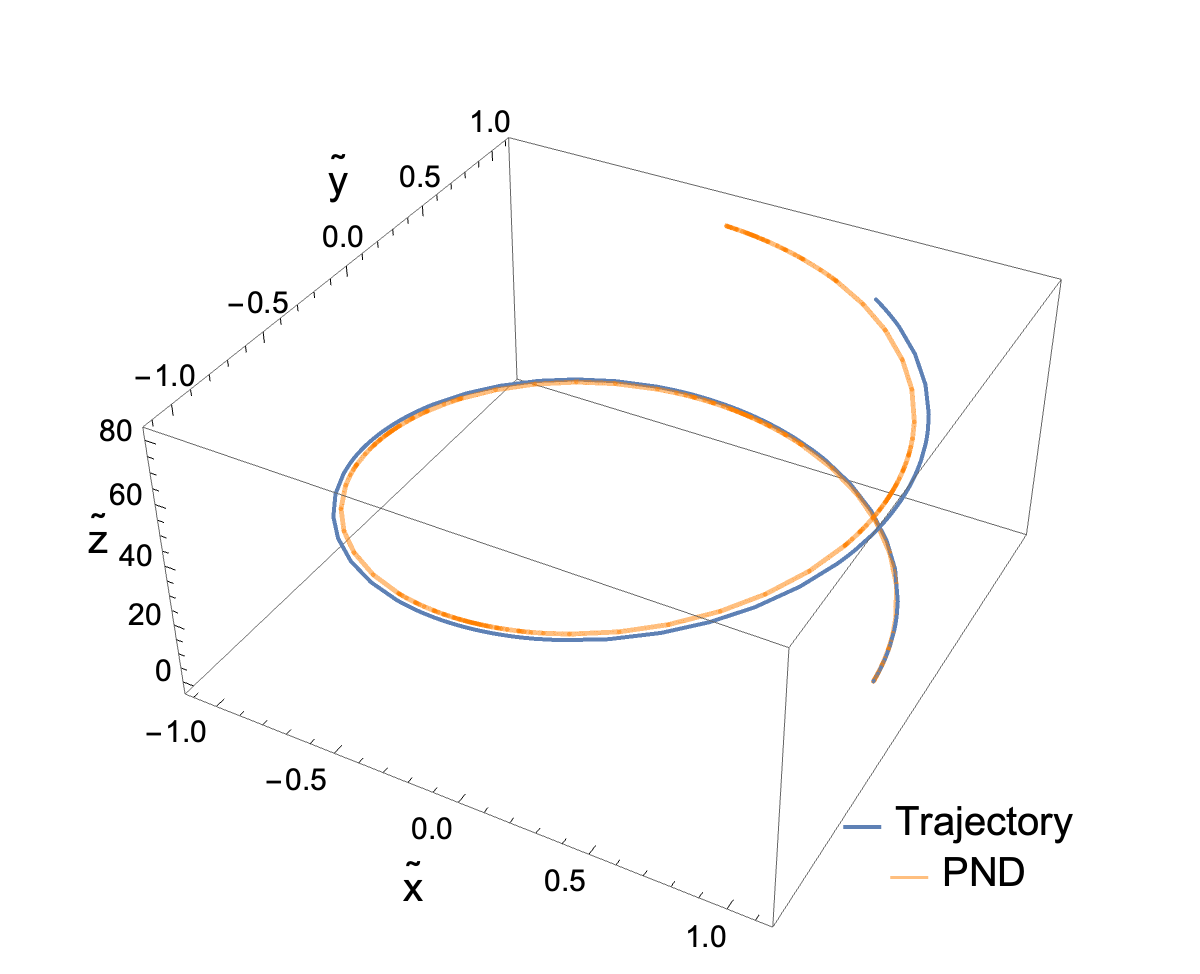}\\
    \caption{Trajectory of a particle under the helical field configuration \eqref{Ehelix} and \eqref{Bhelix} with equal electric and magnetic field strengths.  The particle's trajectory is almost tangent to the PND shown in the plot while drifting outwards slowly.  Notice that the $\tilde{z}$ axis has a compressed scale; the torsion-to-curvature ratio is $\iota/\kappa \approx 10$.
    }
    \label{fig:traj}
\end{figure}

\begin{figure}
    \centering    \includegraphics[width=.45\textwidth]{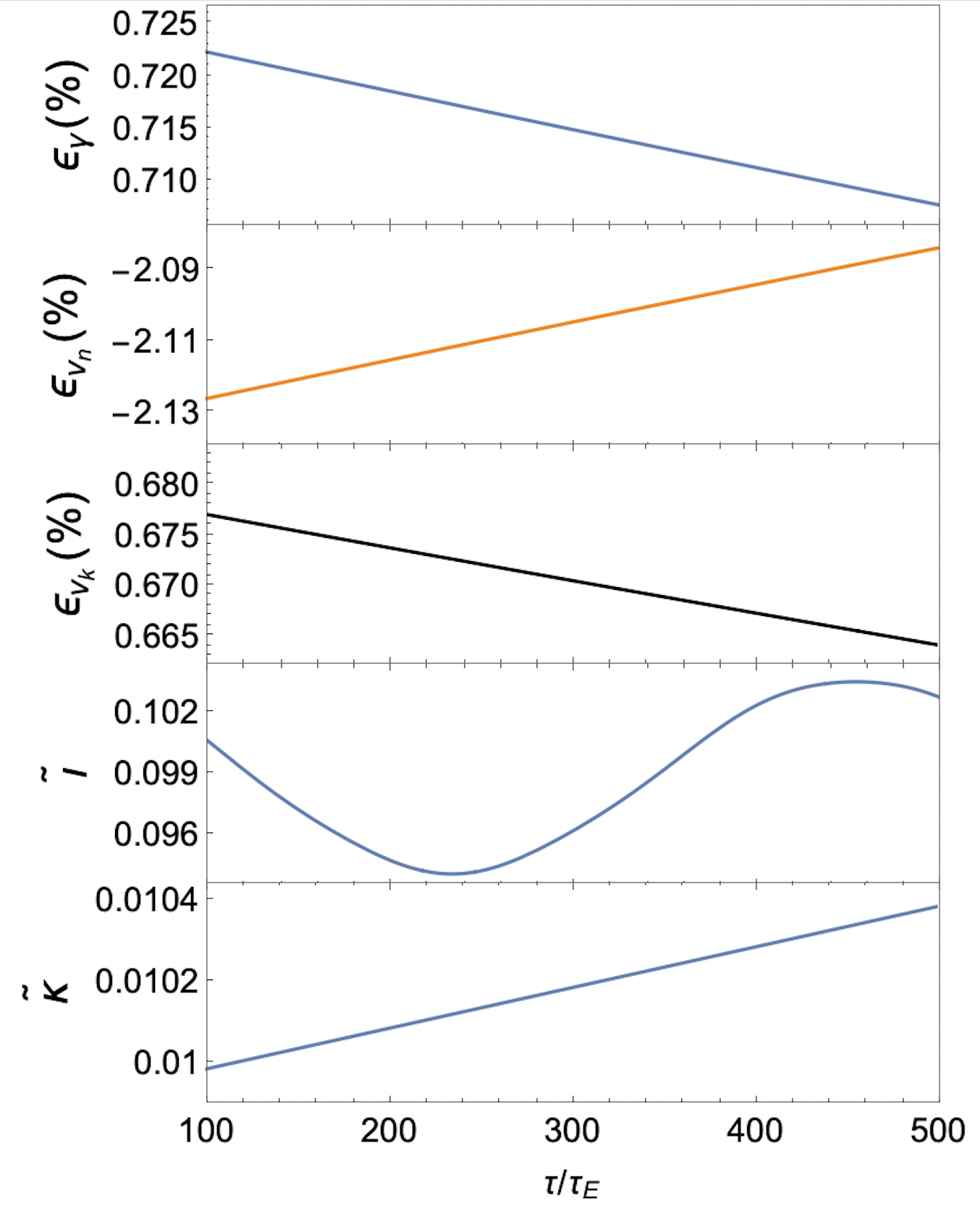}\\
    \caption{Agreement of the analytical predictions \eqref{gammafinal}--\eqref{vkfinal} with numerical simulation in a case with significant torsion.  The trajectory is shown in Fig.~\ref{fig:traj}.  The fractional differences $\epsilon_i$ are defined as the difference between the numerical and analytical value, divided by the analytical value, for $i=\{\gamma,v_n,v_k\}$.  The bottom two plots show the torsion and curvature of the PND at the particle position, normalized as $\tilde{\kappa}=(2/3)\chi^{-2} \mathcal{R} \kappa$ and $\tilde{\iota}=(2/3)\chi^{-2} \mathcal{R} \kappa$ according to the conventions of Sec.~\ref{sec:numerical}.  The $x$-axis uses a timescale $\tau_E=m/(q E_0)$.
    }
    \label{fig:helix}
\end{figure}

In paper I we presented a numerical example of Aristotelian equilibrium in which the torsion was precisely zero.  Here we complement that case with an example where the torsion is significant.  We consider a simple field configuration consisting of parallel electric and magnetic fields tangent to helical curves that fill space.  In terms of some constant $h>0$, the field configuration is
\begin{align}
    \Vec{E} &= \frac{E_0}{\sqrt{h^2+x^2+y^2}}\left\{-y, x, h \right\} \label{Ehelix} \\
    \Vec{B} &= \frac{B_0}{\sqrt{h^2+x^2+y^2}}\left\{-y, x, h \right\}. \label{Bhelix}
\end{align}
The PNDs are also helical, and their curvature and torsion are
\begin{align}
\kappa & = \frac{\sqrt{x^2+y^2}}{x^2+y^2+h^2} \label{kappahelix} \\
\iota & = \frac{h}{x^2+y^2+h^2}, \label{iotahelix}
\end{align}
with ratio
\begin{align}
    \frac{\iota}{\kappa} = \frac{h}{\sqrt{x^2+y^2}}. 
\end{align}
We choose field strengths $\Tilde{E}_0=\tilde{B}_0=1$ and height $\tilde{h}=10$.  For electrons, the physical units of $\tilde{E}_0$ and $\tilde{B}_0$ are given in Eqs.~\eqref{Ephys} and \eqref{Bphys}, while the value of $\tilde{h}$ represents meters.  We start the particle at $(1,0,0)$ and it reaches equilibrium within several timescales $m/(q E)$. 
 Fig.~\ref{fig:traj} shows a portion of the ensuing trajectory, which closely hugs a PND while slowly drifting outwards.  Although the torsion of this PND is times larger than the curvature, we still have $C_3 \approx .02 \ll 1$ and the formulas for the Aristotelian equilibrium should still be valid. Fig.~\ref{fig:helix} shows indeed that the numerical results match the predicted analytical formulas.

\begin{figure*}
     \centering
\includegraphics[width=.85\textwidth]{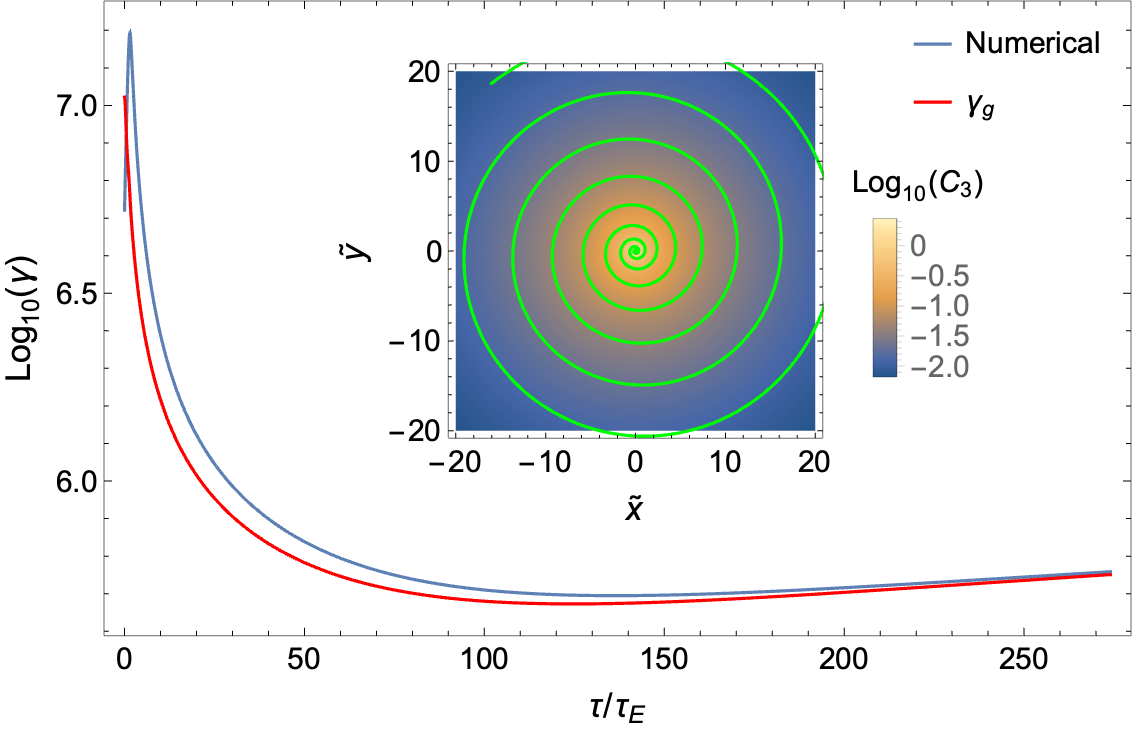}
\caption{An example of evolution through a region where the torsion condition $C_3 \ll 1$ is violated.  Since $C_3\propto \iota/\sqrt{\kappa}$ diverges on the symmetry axis for the nested helical configuration we consider, the torsion condition is violated near the axis.  Here we show an evolution beginning in this region; the particle starts at $\tilde{\vec{x}}=\{0.01,0,0\}$, where $C_3 \approx 5.6$.  The initial velocity is along the PND with initial Lorentz factor equal to half the Gruzinov value.  The particle still approximately follows the PND, but with time-variable Lorentz factor, such that it does not reach equilibrium where $C_3$ is large.  However, it eventually drifts to a region of small $C_3$ and settles down to the Aristotelian equilibrium.\label{fig:spiral}}
 \end{figure*}

 By suitable choice of the parameters $E_0$, $B_0$, and $h$, one can arrange for $C_3$ to be arbitrarily large over an arbitrarily large region of space.  However, since $\kappa$ \eqref{kappahelix} falls off more slowly than $\iota$ \eqref{iotahelix}, the value of $C_3$ always becomes small at large distances from the $z$ axis [see Eq.~\eqref{C3b}].  In numerical experiments we find that equilibrium does not occur in the region of large $C_3$.\footnote{Although $C_3\ll1$ is required for the formulas \eqref{gammafinal}--\eqref{vkfinal} to apply, in principle the particle could reach an equilibrium described by different formulas.  In paper I we in fact derived a more general formula for the Lorentz factor as a quartic equation for $\gamma$ (see Eq.~(61) of paper I) that does not rely on $C_3 \ll 1$.  However, this formula does require the equilibrium assumption \eqref{c4}, and we found in this case that the particle does not enter equilibrium in this sense while $C_3$ is still large.  We have not identified a field configuration where the alternative Lorentz factor formula both applies and is significantly different from the Gruzinov value.} 
 However, the particle still moves primarily along the principal null direction, now with Lorentz factor growing in time, indicating yet another regime of PND motion worthy of future exploration.  In this particular numerical experiment, the motion also has a slower outward drift that eventually takes the particle to a region of small $C_3$ and $C_4$, where its Lorentz factor finally settles down to the Gruzinov value \eqref{gruzigamma}. These properties are illustrated in Fig.~\ref{fig:spiral}
 
 These simple examples demonstrate that the equilibrium works as expected when torsion is non-zero, as long as $C_3 \ll 1$.  In the remainder of the paper we will set the torsion precisely to zero in order to simplify the discussion.

\section{Approach to Equilibrium}\label{sec:approach}

We now study the process by which a particle enters equilibrium, using a combination of analytical and numerical approaches.

\subsection{Uniform field solution}\label{sec:uniform}

At least for some short time, the motion of a particle can be determined in the approximation that the field is uniform.  In this approximation, we can always work in a frame where $\vec{E}$ and $\vec{B}$ are parallel.  The full analytic solution to the LL equation was found in this case by \cite{Heintzmann1973}.  We denote the initial velocity by $(v_1,v_2,v_3)$ with initial Lorentz factor $\gamma_0$, such that the initial four-velocity is given by 
\begin{align}
    u^{\alpha}(0)=\gamma_0 (1,v_1,v_2,v_3).
\end{align}
Taking the field to be in the $z$ direction, the analytic solution to the LL equation is
\begin{align}
    \gamma(\tau)&=\frac{1}{2}\frac{(1+v_3)+(1-v_3)e^{-2\tau/\tau_E}}{\sqrt{1-v_3^2-(v_1^2+v_2^2)e^{-2  \tau/(\delta\tau_E)}}}e^{\tau/\tau_E} \label{uniform_gamma}\\
    v_x(\tau)&=A(\tau)e^{-\tau/\tau_\perp}\sin(\tau/\tau_B+\phi) \label{uniform_vx} \\
    v_y(\tau)&=-\text{sign}(qB_0)A(\tau)e^{-\tau/\tau_\perp}\cos(\tau/\tau_B+\phi) \label{uniform_vy} \\
    v_z(\tau)&=\text{sign}(q)\frac{(1+v_3)-(1-v_3)e^{-2\tau/\tau_E}}{(1+v_3)+(1-v_3)e^{-2\tau/\tau_E}}.  \label{uniform_vz}
\end{align}
We are using Cartesian coordinates with $v_x(\tau=0)=v_1$, $v_y(\tau=0)=v_2$, and $v_z(\tau=0)=v_3$, and $\phi$ is the initial direction of motion in the $xy$ plane ($\tan \phi = v_2/v_1$).  Here $A(\tau)$ is defined by
\begin{align}
    A(\tau)&=\frac{2\sqrt{v_1^2+v_2^2} }{(1+v_3)+(1-v_3)e^{-2\tau/\tau_E}},
\end{align}
and we introduced three time scales
\begin{align}
\tau_{B} & = \frac{m}{|q B_0|}, \label{tauB} \\
\tau_{E} & = \frac{m}{|q| E_0} \label{tauE} \\
\tau_{\perp} & = \frac{m}{|q| E_0}\frac{\delta}{\delta+1}. \label{tauperp}
\end{align}
The definition of $\delta$ was given above in Eq.~\eqref{delta}.

We see that the particle asymptotically approaches light speed in the $z$ direction, i.e., it eventually moves along the PND.  We interpret this to mean that radiation damps only the perpendicular momentum.  The particle executes a damped circular motion in the $xy$ plane (around the field direction), in accordance with the low-$E_0$ intuition of synchrotron motion and associated radiation damping.  The period of oscillation is just the classical synchrotron period $\tau_B$ (expressed here in proper time), while the decay time $\tau_\perp$ generalizes the classical synchrotron damping result.  In particular, we have
\begin{align}
\tau_\perp \approx \begin{cases} \tau_B \frac{\mathcal{E}}{|B_0|} & |B_0| \gg E_0 \\
\tau_E &  E_0 \gg |B_0|
\end{cases},
\end{align}
taking into account $E_0 \ll \mathcal{E}$.  
The former case agrees in order of magnitude with Eq.~(A2) of Ref.~\cite{1984RvMP...56..255B}.

To understand the dynamics it is convenient to expand the Lorentz factor at early and late times,
\begin{align}
    \gamma(\tau) \approx 
    \begin{cases}
    \gamma_0 \left( 1 - \tau/\tau_{\rm drop} - \frac{v_3}{\gamma_0^2} \tau/\tau_E \right), & \tau \to 0 \\
    \frac{1}{2} \sqrt{\frac{1+v^3}{1-v_3^2}} e^{\tau/\tau_E}, & \tau \to \infty.
    \end{cases}\label{gammaearlylate}
\end{align}
where we introduce yet another timescale
\begin{align}\label{taudrop}
    \tau_{\rm drop} \equiv \frac{\delta}{\gamma_0^2(v_1^2+v_2^2)}\tau_E.
\end{align}  
For sufficiently large initial velocity perpendicular to the field, $\gamma_0(v_1^2+v_2^2) \gg \delta$, we have $\tau_{\rm drop} \ll \tau_{\rm E}$.  In this case there is a sudden drop in Lorentz factor on timescale $\tau_{\rm drop}$ as the particle loses its perpendicular momentum, followed by a subsequent rise on timescale $\tau_E$ as the particle gains parallel momentum.  The minimal Lorentz factor is determined by the details of \eqref{uniform_gamma}; in the special case $\delta \gg 1$, the minimum occurs at approximately $\sqrt{\delta}$.

In a realistic field configuration, the exponential rise in Lorentz factor is ultimately cut off by the effects of non-uniform fields.  If the conditions are right for Aristotelian equilibrium, there will be a transition around the equilibrium Lorentz factor.  In order to understand this transition, we now study the LL equation near the Aristotelian equilibrium solution.

\subsection{Near equilibrium solution}\label{sec:near-equilibrium}

To study the dynamics near equilibrium, we adopt the same assumptions as of paper I, assuming time derivatives are perturbatively small (instead of dropping them entirely).  Following the steps of Sec.~IVB of that reference, except retaining time derivatives,\footnote{We also drop all terms involving $\iota$, a step that was justified in a later section of paper I.  When time derivatives are dropped, Eqs.~\eqref{gammatime}, \eqref{vntime}, and \eqref{vktime} reduce, respectively, to Eqs.~ (43), (54) and (55) of paper I with $\iota=0$.} we arrive at three coupled equations for $\gamma$, $v_n$ and $v_k$
\begin{align}
    \frac{d\gamma}{d\tau}&=\frac{|q|\gamma E_0}{m}-\frac{|q|\gamma^3}{m\mathcal{E}}(E_0^2+B_0^2)(v_{n}^2+v_k^2)\label{gammatime} \\
    \frac{d v_n}{d\tau}&= \frac{|q|B_0}{m}v_k-\frac{|q|E_0}{m}\frac{1+\delta}{\delta}v_n-\gamma \kappa \label{vntime} \\
    \frac{d v_k}{d\tau}&=-\frac{|q|B_0}{m}v_n-\frac{|q|E_0}{m}\frac{1+\delta}{\delta}v_k. \label{vktime}
\end{align}
We now express these quantities as small perturbations of their equilibrium values,
\begin{align}
    \gamma&=\Gamma+\bar{\gamma}\\
    v_n&=V_n+\bar{v}_n\\
    v_k&=V_k+\bar{v}_k,
\end{align}
where $\Gamma, V_n, V_k$ are taken to be Eqs.~\eqref{gammafinal}--\eqref{vkfinal}, respectively.  Regarding $E_0,B_0,\kappa$ as constants and linearizing in the barred quantities, we find
\begin{align}
    \tau_E \kappa \frac{d\bar{\gamma}}{d\tau}&=-2 \kappa \bar{\gamma}+2\bar{v}_n/\tau_{\perp}-\epsilon2\bar{v}_k/\tau_B\\
    \frac{d\bar{v}_n}{d\tau}&=-\bar{v}_n/\tau_{\perp}+\epsilon\bar{v}_k/\tau_B-\bar{\gamma}\kappa\\
    \frac{d\bar{v}_k}{d\tau}&=-\epsilon\bar{v}_n/\tau_B-\bar{v}_k/\tau_{\perp},
\end{align}
where the timescales $\tau_i$ were introduced in Eqs.~\eqref{tauB}--\eqref{tauperp} and we have also written $\epsilon = \textrm{sign}(B_0)$.  Defining
\begin{align}
    T & = \frac{\tau}{\tau_E} \\
    b & = \epsilon\frac{\tau_E}{\tau_B} = \frac{B_0}{E_0} \label{b} \\
    c & = \frac{\tau_E}{\tau_\perp} = \frac{1+\delta}{\delta} > 1, \label{c}
\end{align}
this set of equations may also be written
\begin{align}
\frac{d \bar{X}}{dT} = M \bar{X},
\end{align}
with
\begin{align}
    \bar{X} = \begin{pmatrix}
        \bar{\gamma} \kappa \tau_E \\ \bar{v}_n \\ \bar{v}_k 
    \end{pmatrix}, \quad
    M = \begin{pmatrix}
        -2 & 2c & -2b \\
    -1 &-c & b  \\
    0 &-b & -c \\
    \end{pmatrix}.
\end{align}
Making the ansatz
\begin{align}
\bar{X}=\bar{X}_0 e^{\lambda T},
\end{align}
presents the eigenvalue problem $M \bar{X}_0=\lambda \bar{X}_0$ (here $\bar{X}_0$ is a constant independent of $T$).  If $\lambda^{(i)}$ are the eigenvalues and $\bar{X}^{(i)}_0$ the associated eigenvectors (for $i=1,2,3$), then the general solution is
\begin{align}\label{near_equilibrium}
    \bar{X}(T) = \sum_i C_i \textrm{Re}[\bar{X}_0^{(i)} e^{\lambda^{(i)} T}],
\end{align}
for three real constants $C_i$.
The eigenvalues and eigenvectors depend only on $E_0$ and $B_0$ and can be found numerically for any given values of $E_0$ and $B_0$.

We now show that the equilibrium is linearly stable, i.e., $\textrm{Re}[\lambda^{(i)}]<0$.  The eigenvalues $\lambda^{(i)}$ are the roots of the characteristic polynomial (defined with a minus sign for convenience),
\begin{align}\label{flambda}
    &f(\lambda) = - \textrm{det} (M - \lambda I) \nonumber \\ 
    & =\lambda^3+2(c+1)\lambda^2+(b^2+6c+c^2)\lambda+4(b^2+c^2).
\end{align}
Notice that all the coefficients are real and positive.  Thus $f$ is strictly positive for $\lambda \geq 0$, implying that any real root is strictly negative.  We have therefore proven stability in the case of real roots only.  For complex roots, note that such roots must appear in a complex conjugate pair, since we consider a real polynomial.  We may therefore write
\begin{align}\label{complexroots}
    f(\lambda)=(\lambda-\lambda_r)(\lambda-\alpha -i \beta)(\lambda-\alpha + i \beta),
\end{align}
with $\lambda_r,\alpha,\beta$ all real.  Comparing the coefficient of $\lambda^2$ with the polynomial \eqref{flambda}, we find that
\begin{align}\label{oooh}
    2\alpha = -\lambda_r-2(c+1).
\end{align}
However, we find that $f$ is negative at $\lambda=-2(c+1)$:
\begin{align}
    f(-2(c+1))=-2\left[(c+3)(c+2)c+b^2(c-1)\right] <0,
\end{align}
noting from Eq.~\eqref{c} that $c>1$.  Since $f$ is positive for $\lambda \geq 0$, there must be a real root between $\lambda=-2(c+1)$ and $\lambda=0$.  In the case \eqref{complexroots} we consider, $\lambda_r$ is the single real root, so we have
\begin{align}
   -2(c+1) < \lambda_r < 0.
\end{align}
It then follows from \eqref{oooh} that $\alpha$ is strictly negative, completing the proof that $\textrm{Re}[\lambda^{(i)}]<0$. 

Perturbations away from equilibrium are thus exponentially damped.  The qualitative behavior of approach to equilibrium is determined by the mode (or pair of complex conjugate modes) with smallest $|\textrm{Re}[\lambda^{(i)}]|$, which decays the slowest.  If this mode has a non-zero imaginary part (i.e., it is part of a complex-conjugate pair), then oscillations will accompany the decay, with frequency equal to $|\textrm{Im}[\lambda^{(i)}]|$.  Note also that the properties of the decay are insensitive to the sign of $b$, since the characteristic polynomial \eqref{flambda} depends only on $b^2$.

It is interesting to ask for what parameter ranges such oscillations will be important.  The discriminant of the cubic \eqref{flambda} is equal to 
\begin{align}
    \Delta &= -\frac{4}{27}[(3b^2-c^2+10c-4)^3+\\
    &(c^3-15c^2+30c+9b^2c-45b^2-8)^2 ]
\end{align}
When $\Delta < 0$ there are complex-conjugate roots; otherwise all roots are real.  Both signs occur over the parameter ranges $b\in \mathbb{R}, c>1$, so both behaviors are possible.  If $\Delta \geq 0$, there will be no oscillations.  If $\Delta < 0$ there will be oscillations, and these will be the dominant behavior if $\alpha < \lambda_r$ in the notation of \eqref{complexroots}.

One tractable case is when $\delta \gg 1$ so that $c \approx 1$.  It is easy to see that the discriminant is negative for $c=1$, so that there will be complex-conjugate roots.  One can then check that these oscillatory modes decay more slowly than the pure exponential mode provided $|b| \geq 2/\sqrt{3}$, and that they decay with a similar order of magnitude even for $|b| \leq 2/\sqrt{3}$.  The oscillations will thus always be important in the case $\delta \gg 1$.  

Some further details are helpful for interpreting this case. As a function of $b$, the dominant decay time (real part of eigenvalue with real part closest to zero) ranges between $-1$ at $b=0$ and $-4$ at $|b| \to \infty$, and the frequency of oscillations (imaginary part of complex eigenvalue) ranges between (approximately) $1.32$ at $b=0$ and infinity at large $|b|$, approaching linearly like $|b|$ as $b \to \pm\infty$.  Back in the physical time coordinate $\tau= \tau_E T$, we see that decay will typically take several to tens of $\tau_E$, and the oscillation angular frequency will be at least $1.32 \tau_E^{-1}$, and will be approximately $|b|\tau_E^{-1}=\tau_B^{-1}$ at large $|b|$.  It is notable that these oscillations exist even in the pure-E case ($\delta \gg 1, b=0$) and are thus physically distinct from synchrotron motion in that limit.

\begin{figure*}
    \centering
    \includegraphics[width=.85\textwidth]{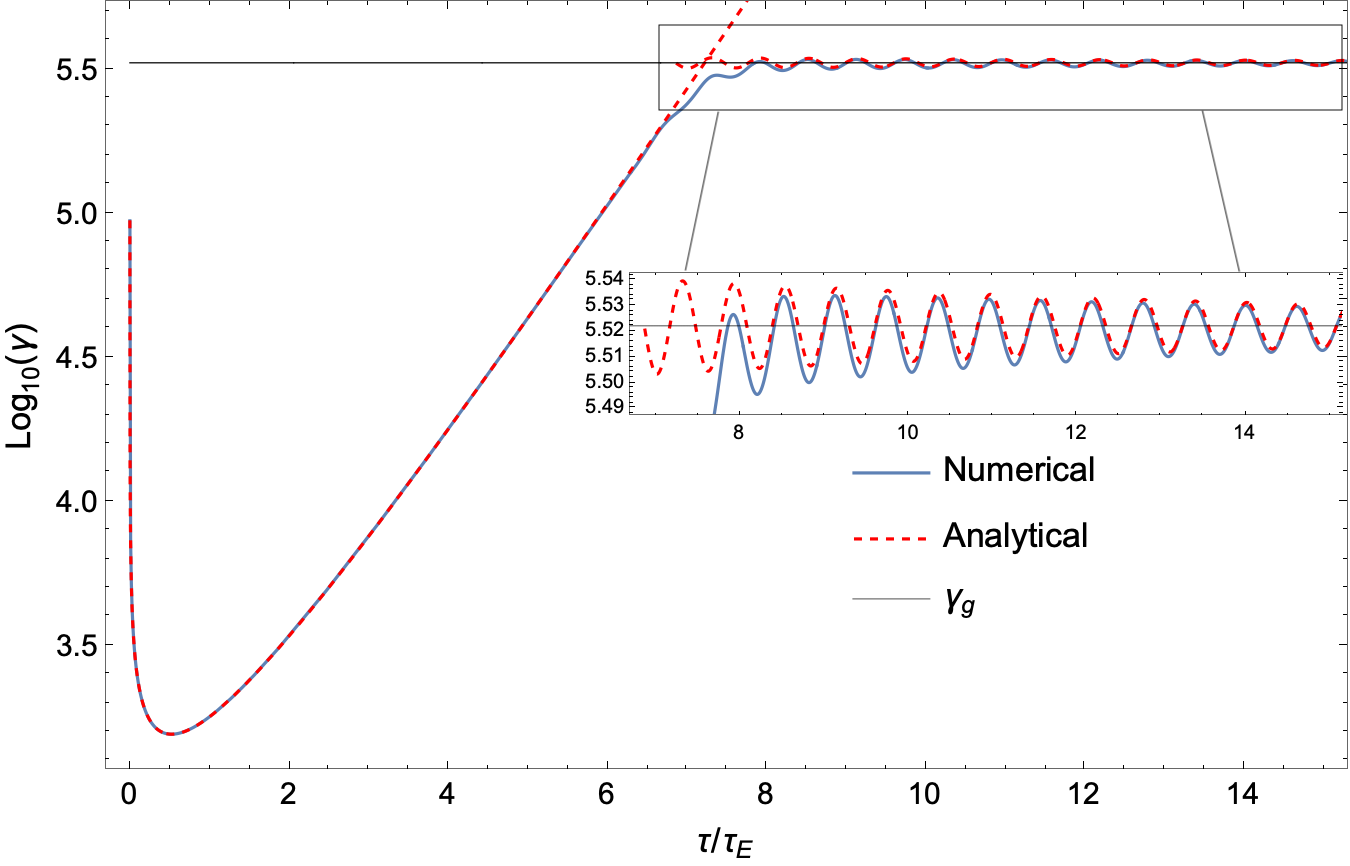}\\
    \caption{An example of entry into equilibrium.  The field configuration is circular with $\tilde{E}_0=1$ and $\tilde{B}_0=10$.  The particle begins at $\tilde{x}=(1,0,0)$ with initial momenta $\tilde{p}=(-2.32,8.28,3.97) \times 10^4$.  The particle quickly loses its perpendicular momentum on timescale $\tau_{\rm drop} \approx 10^{-4} \tau_E$ \eqref{taudrop} before accelerating along the PND to near the equilibrium Lorentz factor over several $\tau_E$ and finally approaching the equilibrium in a weakly damped exponential fashion.  Along with the numerical trajectory, we show the corresponding analytical solutions in the uniform-field and near-equilibrium approximations.  The uniform field solution \eqref{uniform_gamma}--\eqref{uniform_vz} has 5 parameters $E_0, B_0, v_i$, all of which are fixed by the initial data.  The near-equilibrium solution \eqref{near_equilibrium} has six parameters parameters $E_0,B_0,\kappa,C^{(i)}$.  The first three are chosen according to the local field at the initial position and the last three are chosen by fitting with data from $\tau=15\tau_E$ to the end of the evolution at $\tau=30\tau_E$.  The oscillation period is approximately $.61 \tau_E$, and the damping time scale is approximately $10 \tau_E$ (the exponential envelope is $e^{-0.1 \tau/\tau_E}$).  Only the Lorentz factor was used for these fits; however, using the fit parameters, the perpendicular velocities $v_n$ and $v_k$ display a similar level of agreement.
    }
    \label{fig:gammaperp}
\end{figure*}

\subsection{Numerical examples}

We now show two numerical examples illustrating the features explored in the previous subsections.  We consider the case of parallel circular electric and magnetic fields, i.e., Eqs.~\eqref{Ehelix} and \eqref{Bhelix} with $h=0$.  Since the electric and magnetic fields are parallel in the lab frame, no  boost is required to relate the parallel-frame uniform field solution (Sec.~\ref{sec:uniform}) to the lab-frame near-equilibrium solution (Sec.~\ref{sec:near-equilibrium}).  We always place the particle in a region of the circular field configuration where $C_4\ll1$ and hence equilibrium is expected to occur.

First we consider the case where the particle begins with a large momentum perpendicular to the PND.  Based on the analysis of the previous subsections, we expect the particle to sharply lose its perpendicular momentum on a timescale $\tau_{\rm drop}$ \eqref{taudrop}, then gain parallel momentum exponentially on a timescale $\tau_E$ \eqref{tauE} until it nears the equilibrium Lorentz factor, and finally approach equilibrium exponentially, with timescale and possible oscillations determined from $\tau_E,\tau_B,\tau_\perp$ via the eigenvalue problem discussed in Sec.~\ref{sec:near-equilibrium}.  These expectations are confirmed by our numerical experiments; an example is shown in Fig.~\ref{fig:gammaperp}.

\begin{figure}
    \centering
    \includegraphics[width=.45\textwidth]{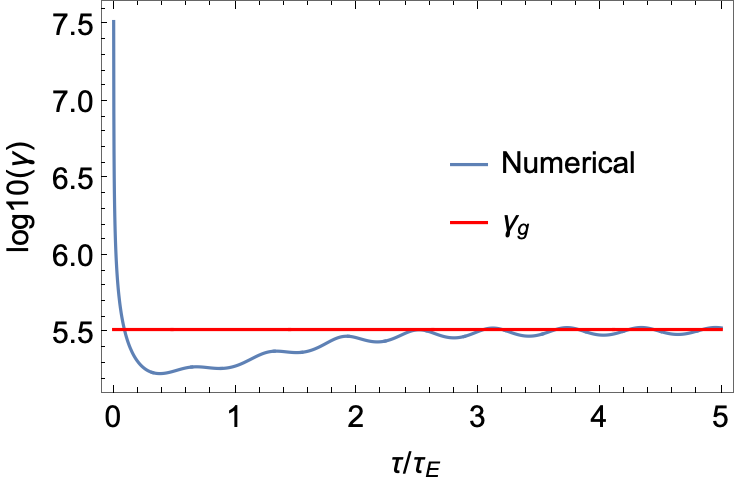}\\
    \caption{Entry into equilibrium after starting with large momentum parallel to the PND.  The field setup is the same as Fig.~\ref{fig:gammaperp}, except that the particle starts with momentum entirely in the $y$ direction with Lorentz factor equal to 100 times the equilibrium value, $\gamma_0=100\gamma_{g}$.  The particle quickly loses energy until near the equilibrium value and then approaches equilibrium with a sinusoidally modulated exponential.
    }
    \label{fig:gammapar}
\end{figure}

We next consider the case where the particle begins with a large momentum $\gamma \gg \gamma_g$ parallel to the PND.  Here the uniform field approximation is not  useful since the particle quickly reaches the region where the field line bends away from its motion.  However, at this stage we can regard the particle as traveling through a new uniform field configuration with some perpendicular momentum, which will be removed by radiation reaction according to the intuition discussed in the first new paragraph below Eq.~\eqref{tauperp}.  In this way the particle will lose energy until it either enters equilibrium directly (``from above'') or finds itself in a uniform field configuration that will accelerate it back up to near-equilibrium conditions, after which it enters equilibrium ``from below''.  An example somewhat intermediate between these two cases is shown in Fig.~\ref{fig:gammapar}.

\section{Numerical survey}\label{sec:C4}

Up until now, we have explored the properties of the Aristotelian equilibrium and the manner in which particles enter the equilibrium.  We now turn to the general question of when particles will indeed enter the equilibrium.  As reviewed in Sec.~\ref{sec:conditions}, paper I identified five necessary conditions $C_i \ll 1$ for equilibrium to occur.  We now provide numerical evidence that these conditions are indeed sufficient for it to occur, and we use the results to gain a more quantitative understanding of how well they must be satisfied.

Exploring all five conditions would be computationally intractable.  However, given that $C_4 \ll 1$ implies $C_1 \ll 1$ and $C_2 \ll 1$ outside of extremely finely tuned field configurations (Eqs.~\eqref{ic41} and \eqref{ic42}), we can safely ignore $C_1$ and $C_2$.  The case where $C_3 \ll 1$ is violated is potentially interesting for helical fields (as occur, e.g., in relativistic jets), and we found that equilibrium does not occur in this case (Sec.~\ref{sec:torsion}).  Violations of the quasi-static assumption $C_5 \ll 1$ are certainly interesting (e.g., for laser fields), but outside the scope of this paper.  Instead, we will pick field configurations where $C_3=0$ and $C_5=0$ (exactly torsion-free and static, respectively) and explore the role of $C_4$ in determining whether equilibrium occurs.

\subsection{Circular fields and the role of $\eta$}\label{sec:circular}

Eq.~\eqref{eta} defined a quantity $\eta$ that appears in the conditions for equilibrium.  For generic fields, $\eta \approx 1$ and factors of $\eta$ can be dropped.  To illustrate a situation where $\eta$ \textit{cannot} be dropped, consider a purely azimuthal field configuration with constant parallel electric and magnetic fields.  The field lines and PNDs are just circles of radius $\rho=\sqrt{x^2+y^2}$.  Since $\vec{\nabla}\rho=\vec{n}$, where $\vec{n}$ is the Frenet-Serret normal direction to the curve, $\eta$ can be calculated as
\begin{align}\label{eta4_circular}
    \eta = |\Vec{v} \cdot \nabla \rho|= |v_n|.
\end{align}
This quantity is small in equilibrium since $|v_n| \ll 1$ by the assumption of motion along a PND.  The small size of $\eta$ arises because the field strength and radius of curvature are both constant along the PND direction $\vec{\ell}$, a finely tuned arrangement.  From Eqs.~\eqref{vnfinal} and \eqref{gammafinal}, we have
\begin{align}
    |v_n|^2=\frac{2}{3}\frac{\mathcal{R}}{R}\sqrt{\frac{E_0}{\mathcal{E}}}(1+\delta)^2.
\end{align}
The condition \eqref{C4b} then becomes 
\begin{align}\label{circular_validity}
    R^2 E_0 \gg (1+\delta)^2 \mathcal{R}^2\mathcal{E},
\end{align}
where we drop a factor of $2/3$. 

Eq.~\eqref{circular_validity} is a necessary condition for equilibrium to occur.  To check whether it is also sufficient, we simulated a large number of trajectories with different field configuration parameters and particle initial conditions.  Specifically, we fixed $\tilde{B}_0=0.1$ and chose the particle to begin on the $x$ axis, choosing the other parameters randomly from uniform distributions in the ranges $\log_{10} \tilde{E}_0 \in (-4,0)$ for the log of the field strength,  $\gamma \in (1,\gamma_g)$ for the initial Lorentz factor, $\theta\in(0,\pi)$ and $\phi\in(0,2\pi)$ for the initial direction of motion (where $\theta,\phi$ are spherical coordinates on the space of velocities), and $\log_{10} x \in (-3,0)$ for the log of the initial position.  The initial radius of curvature $\tilde{R}$ is just $x$, so we sample an initial range $\log_{10} \tilde{R} \in (-3,0)$. 

\begin{figure}
    \centering
    \includegraphics[width=.45\textwidth]{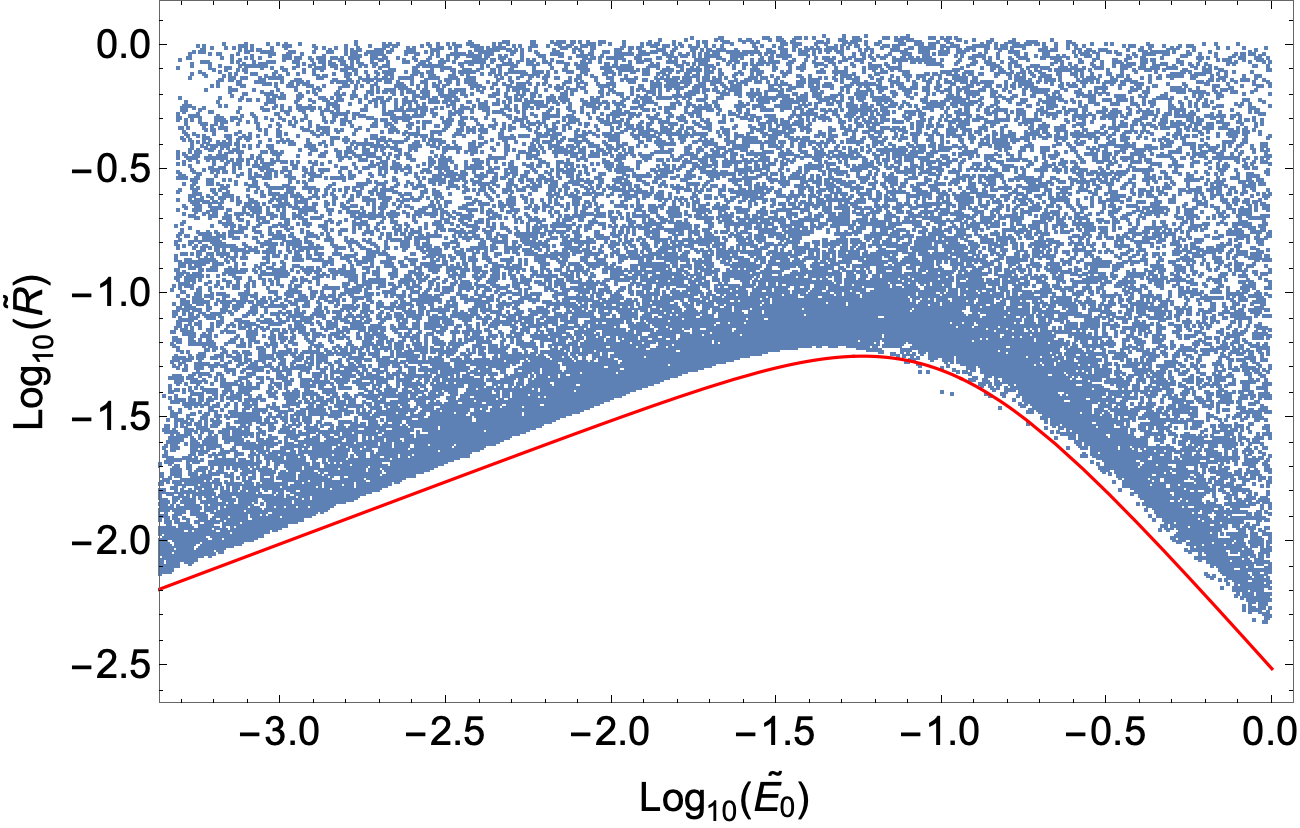}\\
    \caption{Values of the local electric field and PND curvature radius at entry into equilibrium in a large parameter survey.  In this circular field configuration, equilibrium is expected to occur when $N=R^2 E_0/[(1+\delta)^2 \mathcal{R}^2 \mathcal{E}]\gg1$ [Eq.~\eqref{circular_validity}].  The numerics validate this condition: the red line is the curve $N=15$.
    }
    \label{fig:CondPlot_circular}
\end{figure}

For each run of the code we evolve for a maximum time of $\tau_{\rm max}=(6 \log \gamma_g )\tau_E$, which is sufficient for the particle to enter equilibrium if it is destined to do so.\footnote{According to Eq.~\eqref{gammaearlylate}, the time for a particle to be accelerated to $\gamma_g$ is of order $\tau_E \log \gamma_g$.  For the circular field $E_0$ and hence $\tau_E$ is constant; however, for other field configurations we update the maximum allowed run time after each time step to use the local value of $E_0$.  We also examined a selection of the trajectories that did not enter equilibrium in this time and evolved them for longer, finding that indeed they never enter equilibrium.}  At any given time in the trajectory, the particle is considered to have entered equilibrium if $\langle |(\gamma-\gamma_g)|/\gamma_g\rangle<3\%$, where the average $\langle ... \rangle$ is calculated over the last $20\%$ of the computed trajectory, using the local value of $\gamma_g$.  Once a particle enters equilibrium, we record the local electric field $E_0$ and curvature radius $R$ and terminate the run.  (If the particle never enters equilibrium, we make no corresponding record.)  Fig.~\ref{fig:CondPlot_circular} shows the values of $E$ and $R$ at which equilibrium occurred in our random sample, showing clearly that the condition \eqref{circular_validity} indeed controls whether equilibrium is obtained.  We find that the left-hand-side must be approximately 15 times larger than the right-hand side (red line in the figure) for equilibrium to occur.

When $E_0 \gg B_0$, we have $\delta \approx \mathcal{E}/E_0 \gg 1$, and this condition reduces to the original Gruzinov condition \eqref{gruzicond}.  (This agreement is accidental, due to the specific form of $\eta$ for this field configuration.)  By contrast, when $E_0 \lesssim B_0$, the condition \eqref{circular_validity} remains distinct from the Gruzinov condition \eqref{gruzicond}.  This behavior is seen clearly in Fig.~\ref{fig:CondPlot_circular}.  The Gruzinov condition is necessary, but not sufficient, in this special case.  The correct condition for entry into equilibrium is the modified condition \eqref{C4b}.

\subsection{More general field configurations}

The circular field is a special case that cleanly illustrates the role of $\eta$ in the condition $C_4\ll1$ \eqref{C4} for entry into equilibrium.  In order to test the condition more generally, we consider two more kinds of field configuration.  The ``ellipse'' field configuration simply changes the shape of the field lines to ellipses instead of circles, while keeping everything else the same.  The ``separate center'' field configuration consists of circular electric and magnetic field lines with equal field strengths $E_0=B_0$ but with the center of the circles shifted by a distance $d$.  We randomly varied the field strength and center-distance while also randomly choosing initial conditions as before.  

\begin{figure}
    \centering
    \includegraphics[width=.45\textwidth]{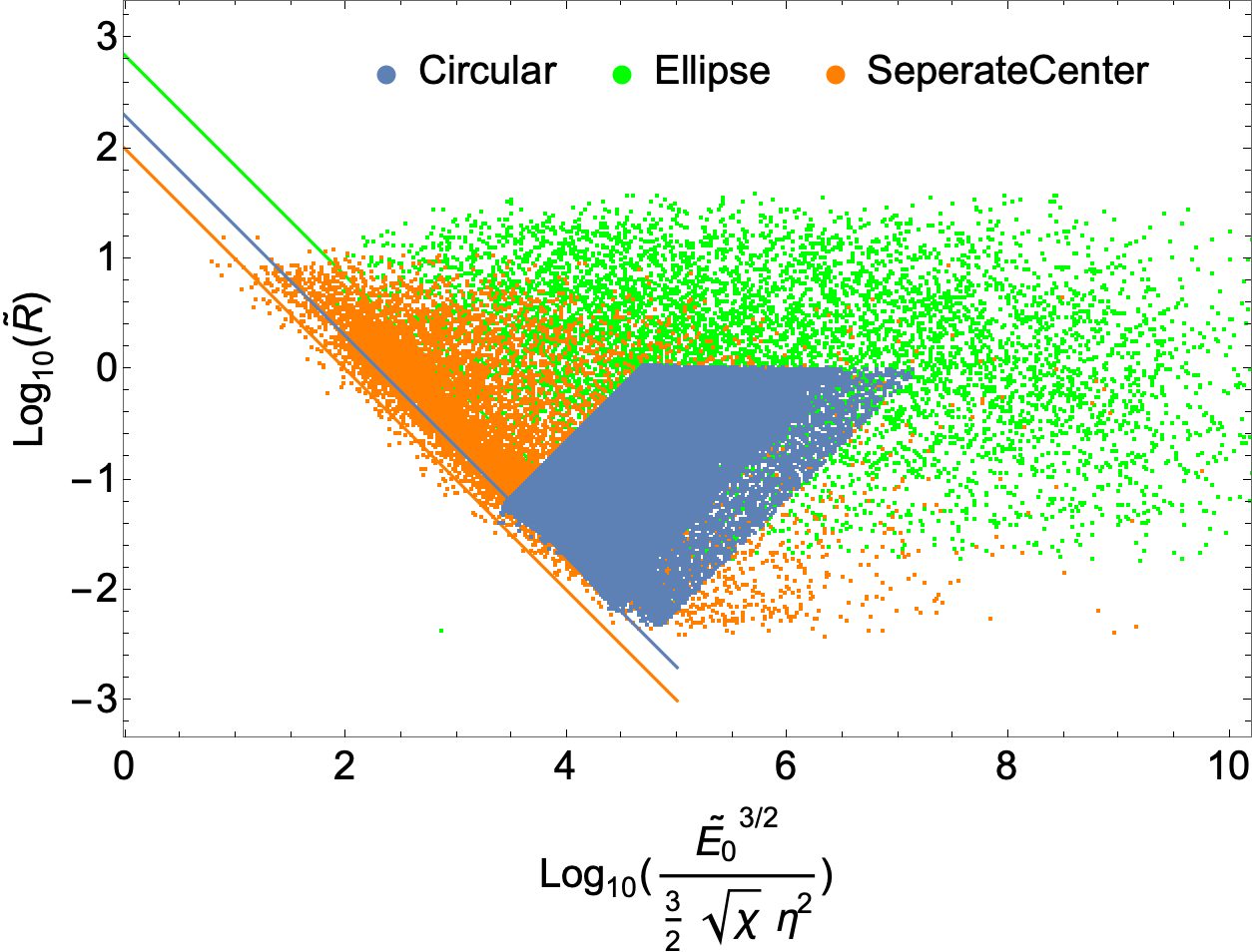}\\
    \caption{Validity of the main condition for entry into equilibrium [$C_4\ll1$, \eqref{C4} or \eqref{C4b}] in three numerical parameter surveys.  The colored dots correspond to the local field properties where a particle in the corresponding survey entered equilibrium.  Lines of constant $C_4$ have slope $-1$ on this plot and the $y$-intercept is the log of $1/C_4^2$.  For each field configuration, the edge of the region of equilibria clearly has the predicted slope of $-1$.  We have drawn approximate reference lines for this edge to guide help guide the eye.  Entry to equilibrium occurs when $C_4 \lesssim 5\%$.}
    \label{fig:CondPlot_all}
\end{figure}

The results of these surveys are shown along with the circular case in Fig.~\ref{fig:CondPlot_all}, demonstrating clearly that the condition $C_4 \ll 1$ is necessary and sufficient for equilibrium to occur.  The precise value of $C_4$ required depends on the field configuration, ranging from $\sim0.1$ to $\sim0.01$ for the cases considered here.

\section{Summary of results}\label{sec:summary}

We have provided a detailed understanding of the entry into Aristotelian equilibrium, identifying the relevant timescales, giving analytical descriptions where possible, and showing that numerical simulations match the predicted behavior.  We showed analytically that the equilibrium is linearly stable and identified the presence of oscillations at entry for a large region of parameter space.  We performed numerical parameter surveys exploring the conditions for equilibrium to occur.  Combined with the results of paper I, this study provides strong evidence that the conditions \eqref{C1}--\eqref{C5} are necessary and sufficient for Eqs.~\eqref{gammafinal}--\eqref{vkfinal} to describe the motion of a charged particle.  This provides a solid foundation for using the Aristotelian approximation in astrophysical modeling.

\section*{Acknowledgements} This work was
supported in part by NSF grant PHY–1752809, and NSF Grants
PHY-1912619 and PHY-2145421, to the University of Arizona.

\appendix

\section{Numerical Method}\label{sec:numerical-details}

Our numerical scheme is based on the implicit 4th order Runge-Kutta-Nystr\"om method.  We first review this method before presenting an improved iteration method and an adaptive timestep version that we also used in some cases.

For a second-order ordinary differential equation
\begin{align}
\Ddot{x}^{\alpha}=f^{\alpha}(t,x^{\alpha},\Dot{x}^{\alpha}),
\end{align}
the implicit 4th order Runge-Kutta-Nystr\"om method can be expressed as follows.  First fix a time step $h$.  If $x_n^{\alpha}$ and $\Dot{x}_n^{\alpha}$ represent the value and derivative at time $t_n$, then the values at the next step $t_{n+1}=t_n+h$ are obtained by
\begin{align}
x_{n+1}^{\alpha}&=x_{n}^{\alpha}+h\Dot{x}_n^{\alpha}+h^2 b_i k^{\alpha}_i\\
    \Dot{x}_{n+1}^{\alpha}&=\Dot{x}_n^{\alpha}+h a_i k^{\alpha}_i,
\end{align}
where $a_i$ and $b_i$ are
\begin{align}
    a_i&=\left(\frac{1}{2},\ \frac{1}{2}\right)\\
    b_i&=\left(\frac{1}{4}+\frac{\sqrt{3}}{12},\ \frac{1}{4}-\frac{\sqrt{3}}{12}\right),
\end{align}
and $k^{\alpha}_i$ must satisfy
\begin{align}\label{k}
    k^{\alpha}_i=f^{\alpha}&\big(t_n+c_i h, x^{\alpha}_n+c_i h \Dot{x}_n^{\alpha}+h^2B_{ij}k^{\alpha}_j, 
    \Dot{x}_n^{\alpha}+hA_{ij}k^{\alpha}_j \big).
\end{align}
In these expressions, repeated indices are summed. Here $c_i$ is the vector
\begin{align}
    c_i=\begin{pmatrix} \frac{1}{2}-\frac{\sqrt{3}}{6} \\ \frac{1}{2}+\frac{\sqrt{3}}{6}\end{pmatrix}
\end{align}
and $A_{ij}$ and $B_{ij}$ are the matrices
\begin{align}
    A_{ij}& =\begin{pmatrix}
    \frac{1}{4} & \frac{1}{4}-\frac{\sqrt{3}}{6}\\
    \frac{1}{4}+\frac{\sqrt{3}}{6} & \frac{1}{4}
    \end{pmatrix}
     \\
    B_{ij}& =\begin{pmatrix}
    \frac{1}{36} & \frac{5}{36}-\frac{\sqrt{3}}{12}\\
    \frac{5}{36}+\frac{\sqrt{3}}{12} & \frac{1}{36}
    \end{pmatrix}.
\end{align}
\subsection*{Iteration method}
Eq.~\eqref{k} can be solved by a fixed point iteration method as follows.  The initial values of $k^{\alpha}_i$ are taken to be
\begin{align}
    (k^{\alpha}_i)^0=f^{\alpha}(t_n, x^{\alpha}_n, \Dot{x}_n^{\alpha}) \ \ (i=1,2).
\end{align}
We then iteratively improve the guess by calculating
\begin{align}\label{iter_k}
    (k^{\alpha}_i)^{N+1}=f^{\alpha}&\big(t_n+c_i h, \nonumber \\ & x^{\alpha}_n+c_i h \Dot{x}_n^{\alpha}+h^2 B_{ij}(k^{\alpha}_j)^{N} \nonumber \\ &
    \Dot{x}_n^{\alpha}+A_{ij}(k^{\alpha}_j)^{N} \big).
\end{align}

However, for large Lorentz factors we found that this iteration method sometimes converged very slowly or got stuck in a limit cycle.  In these cases we instead used overrelaxation with the secant method,
\begin{align}
    y_n& = w\left[y_{n-1}-F(y_{n-1})\frac{y_{n-1}-y_{n-2}}{F(y_{n-1})-F(y_{n-2})}\right] \\ & +(1-w)y_{n-1},
\end{align}
where $y_N=(k_i^{\alpha})^{N}$, $F^{\alpha}_i(k_i^{\alpha})=f^{\alpha}(k^{\alpha}_i)-k_i^{\alpha}$ and $w \in [0,1]$ is the weight.  We chose $w=0.05$.  At each time step we try the simple iteration method first, jumping to the Secant method only if the simple iteration fails to reach a given tolerance (we chose fractional error of $10^{-10}$) within a fixed number of iterations (we chose $10^3$).

\subsection{Adaptive time step}

For most portions of a given evolution the main timescales are $\tau_E$ and $\tau_B$, defined in Eqs.~\eqref{tauE} and \eqref{tauB}.  However, for some initial conditions (e.g., Fig.~\ref{fig:gammaperp}) there can be abrupt changes in Lorentz factor on the much smaller timescale $\tau_{\rm drop}$ defined in Eq.~\eqref{taudrop}.  In most cases we use an adaptive time step scheme, beginning with $h=\rm{min}(\tau_B,\tau_E)\times 1\%$ and updating as follows.

We focus on the Lorentz factor as a simple scalar representative of the solution. Suppose we are at time $t$ in the evolution and consider the value at time $t+2h$, where $h$ is the current step size.  We can compute this either by taking a single step of size $2h$ or by taking two steps of size $h$.  We will denote the resulting values for $\gamma$ by $\gamma_{\rm num}^{(2h)}$ and $\gamma_{\rm num}^{(h)}$, respectively.  Since our method has fourth-order accuracy, these are related to the true value $\gamma_{\rm true}$ by
\begin{align}
    \gamma_{\rm num}^{(h)} & \approx \gamma_{\rm true}(t+2h) + \phi h^4 \\
    \gamma_{\rm num}^{(2h)} & \approx \gamma_{\rm true}(t+2h) + 16 \phi h^4,
\end{align}
where $\phi$ is some unknown constant. We can estimate $\phi$ by solving,
\begin{align}\label{phi}
    \phi \approx \frac{ \gamma_{\rm num}^{(2h)}-\gamma_{\rm num}^{(h)}}{15 h^4}.
\end{align}

Suppose we instead consider a new time step $h_0$ and evolve forward one step to $\gamma^{(h_0)}_{\rm num}$. 
 Now we have 
\begin{align}
\gamma^{(h_0)}_{\rm num}\approx \gamma_{\rm true}(t+2h) + \phi h_0^4,
\end{align}
If we want to achieve an accuracy of $\epsilon = \left|(\gamma_{\rm num}^{(h_0)}-\gamma_{\rm true})/\gamma_{\rm true}\right|$, the time step we should use satisfies
\begin{align}
h_0^4 = \frac{\epsilon \gamma_{\rm num}^{(h)} }{|\phi|}.
\end{align}
Using the formula \eqref{phi} for $\phi$, we conclude that an appropriate time step is
\begin{align}\label{h0}
    h_0 = \zeta h\left(\frac{15\epsilon \gamma_{\rm num}^{(h)} }{|\gamma_{\rm num}^{(2h)}-\gamma_{\rm num}^{(h)}|}\right)^{1/4}.
\end{align}
The factor of $\zeta<1$ is to guarantee that $h_0$ is smaller than that which is expected to exactly achieve the desired error tolerance $\epsilon$, which corresponds to $\zeta=1$. We choose $\zeta=(14/15)^{1/4}$ in our code, and $\epsilon=10^{-6}$.

At each step in the evolution we calculate the candidate new time step $h_0$ according to \eqref{h0}.  Before adopting this as the new time step we consider two potential adjustments.  First, we ensure that $h$ does not change by more than a factor of two.  That is, if $h_0$ is larger than $2h$, then we use $2h$ instead; similarly, if $h_0$ is smaller than $h/2$, we use $h/2$ instead.  Finally, we ensure that the candidate new time step satisfies the condition for the convergence of fixed point iteration, namely that the spectral radius of the (six-dimensional) Jacobian matrix of $f^\alpha_i(k^\beta_j)$ is strictly less than $1$, i.e., the maximum of the absolute values of the eigenvalues of this matrix is less than 1.  (Here $f^\alpha_i$ denotes the RHS of \eqref{k}.) If this test fails, then we divide the step size in half and try again, iterating until a step size satisfying the condition has been found.

\begin{figure}[h]
    \centering
    \includegraphics[width=.45\textwidth]{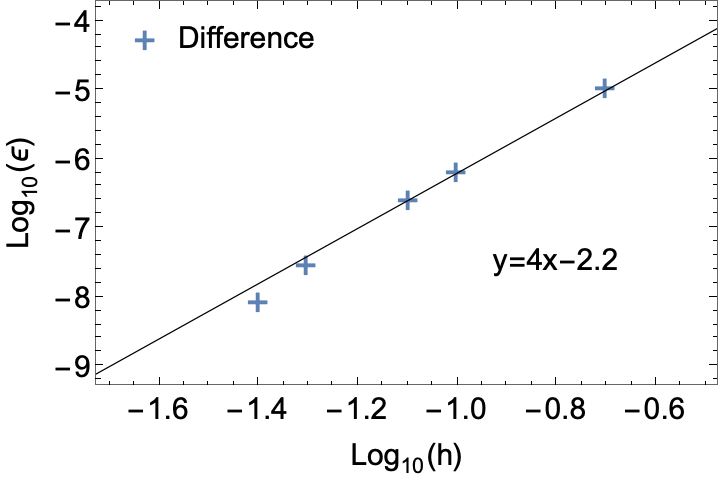}\\
    \caption{Test demonstrating our code's error convergence at 4th order. The black line represents the linear function with slope of 4; the equation given in the plot.
    }
    \label{fig:convergence}
\end{figure}

\subsection{Convergence test}

We performed a convergence test for the code using the uniform field setup, because it has exact analytical solutions. We choose the field strengths as $\tilde{E}_0=0.1,\tilde{B}_0=1$. The initial value we used is $\tilde{\vec{p_0}}=\{100,400,-300\}$. The Lorentz factor Eq.~\eqref{uniform_gamma} at $\tau= 2 \tau_E$ was used as the reference value. The code was executed at different time steps $h$. The fractional difference:
\begin{align}
     \epsilon=\left|\frac{\gamma_{\rm numeric}-\gamma_{\rm analytic}}{\gamma_{\rm analytic}}\right|    
\end{align}
was then calculated. The result of the convergence test is shown in Fig.~\ref{fig:convergence}. It is clear that our code exhibits 4th-order convergence.

\bibliographystyle{utphys}
\bibliography{refs}

\providecommand{\noopsort}[1]{}\providecommand{\singleletter}[1]{#1}%
\providecommand{\href}[2]{#2}\begingroup\raggedright\begin{thebibliography}{10}

\bibitem{synge1956relativity}
J.~L. Synge, {\em Relativity: the special theory}.
\newblock North-Holland Publishing Company Amsterdam, 1956.

\bibitem{mestel1985}
L.~Mestel, J.~Robertson, Y.-M. Wang, and K.~Westfold, ``The axisymmetric pulsar
  magnetosphere,'' {\em Monthly Notices of the Royal Astronomical Society} {\bf
  217} (1985), no.~3, 443--484.

\bibitem{BFink1989}
B.Finkbeiner, H.Herold, T.~Ertl, and H.~Ruder {\em Astronomy and astrophysics}
  {\bf 225} (1989) 479.

\bibitem{gruzinov2013b}
A.~{Gruzinov}, ``{Pulsar Emission Spectrum},'' {\em arXiv e-prints} (Sept.,
  2013) arXiv:1309.6974, \href{http://www.arXiv.org/abs/1309.6974}{{\tt
  1309.6974}}.

\bibitem{Jacobson:2015cia}
T.~Jacobson, ``{Structure of Aristotelian Electrodynamics},'' {\em Phys. Rev.
  D} {\bf 92} (2015), no.~2, 025029,
  \href{http://www.arXiv.org/abs/1504.07311}{{\tt 1504.07311}}.

\bibitem{Petri:2019tix}
J.~P\'etri, ``{Pulsar gamma-ray emission in the radiation reaction regime},''
  {\em Mon. Not. Roy. Astron. Soc.} {\bf 484} (2019), no.~4, 5669--5691,
  \href{http://www.arXiv.org/abs/1901.11439}{{\tt 1901.11439}}.

\bibitem{Cao:2019uhv}
G.~Cao and X.~Yang, ``{Three-dimensional dissipative pulsar magnetospheres with
  Aristotelian electrodynamics},''
  \href{http://www.arXiv.org/abs/1912.00335}{{\tt 1912.00335}}.

\bibitem{Gonoskov:2017lyz}
A.~Gonoskov and M.~Marklund, ``{Radiation-dominated particle and plasma
  dynamics},'' {\em Phys. Plasmas} {\bf 25} (2018), no.~9, 093109,
  \href{http://www.arXiv.org/abs/1707.05749}{{\tt 1707.05749}}.

\bibitem{Samsonov:2018skj}
A.~S. Samsonov, E.~N. Nerush, and I.~Y. Kostyukov, ``{Asymptotic electron
  motion in the strongly-radiation-dominated regime},'' {\em Phys. Rev. A} {\bf
  98} (2018), no.~5, 053858, \href{http://www.arXiv.org/abs/1807.04071}{{\tt
  1807.04071}}.

\bibitem{Ekman:2021vwg}
R.~Ekman, T.~Heinzl, and A.~Ilderton, ``{Exact solutions in radiation reaction
  and the radiation-free direction},'' {\em New J. Phys.} {\bf 23} (2021),
  no.~5, 055001, \href{http://www.arXiv.org/abs/2102.11843}{{\tt 2102.11843}}.

\bibitem{Gonoskov:2021hwf}
A.~Gonoskov, T.~G. Blackburn, M.~Marklund, and S.~S. Bulanov, ``{Charged
  particle motion and radiation in strong electromagnetic fields},''
  \href{http://www.arXiv.org/abs/2107.02161}{{\tt 2107.02161}}.

\bibitem{Samsonov2022}
E.~A.S.Samsonov and I.Yu.Kostyukov, ``{High-order corrections to the
  radiationfree dynamics of an electron in the strongly radiation-dominated
  regime},'' {\em Matter and Radiation at Extremes} {\bf 8} (2022), no.~1,
  014402, \href{http://www.arXiv.org/abs/2208.00673}{{\tt 2208.00673}}.

\bibitem{Cai:2022mkw}
Y.~Cai, S.~E. Gralla, and V.~Paschalidis, ``{Dynamics of ultrarelativistic
  charged particles with strong radiation reaction. I. Aristotelian equilibrium
  state},'' \href{http://www.arXiv.org/abs/2209.07469}{{\tt 2209.07469}}.

\bibitem{LL}
L.~D. Landau and E.~M. Lifshitz, {\em The Classical Theory of Fields: Volume 2
  (Course of Theoretical Physics Series) 4th Edition}.
\newblock Butterworth-Heinemann, 1980.

\bibitem{Gruzinov:2013pva}
A.~Gruzinov, ``{Aristotelian Electrodynamics solves the Pulsar: Lower
  Efficiency of Strong Pulsars},''
  \href{http://www.arXiv.org/abs/1303.4094}{{\tt 1303.4094}}.

\bibitem{Heintzmann1973}
H.~Heintzmann and E.~Schrüfer, ``{Exact solutions of the Lorentz-Dirac
  equations of motion for charged particles in constant electromagnetic
  fields},'' {\em Phys. Rev. A} {\bf 43} (1973), no.~3, 287--288.

\bibitem{1984RvMP...56..255B}
M.~C. {Begelman}, R.~D. {Blandford}, and M.~J. {Rees}, ``{Theory of
  extragalactic radio sources},'' {\em Reviews of Modern Physics} {\bf 56}
  (Apr., 1984) 255--351.

\end{thebibliography}\endgroup

\end{document}